\documentstyle[epsf,fleqn,espcrc2]{article}

\ifx\DeclareFontShape\undefined 
    \newcommand{\mathbf}[1]{{\bf #1}} 
    \newcommand{\mathrm}[1]{{\rm #1}} 
    \newcommand{\mathcal}[1]{{\cal #1}} 
    \newcommand{\mathsf}[1]{{\sf #1}} 
\else 
    \DeclareSymbolFont{lasy}{U}{lasy}{m}{n}
    \SetSymbolFont{lasy}{bold}{U}{lasy}{b}{n}
    \let\Box\undefined
    \DeclareMathSymbol\Box{0}{lasy}{"32}
\fi 
\newcommand{\AmS}{{\protect\the\textfont2
  A\kern-.1667em\lower.5ex\hbox{M}\kern-.125emS}}
\newcommand{\pspicture}[1]{%
\centerline{\setlength\epsfxsize{7.25cm}\epsffile{#1}}}

\newcommand{\figurebox}[2]{\fbox{\vbox to#2in{\hbox to #1in{\hfil} \vfil}}}
\newcommand{\beq}{\begin{equation}}
\newcommand{\eeq}{\end{equation}}
\newcommand{\beqn}{\begin{eqnarray}}
\newcommand{\eeqn}{\end{eqnarray}}

\newcommand{\BtoK}{B \rightarrow K^* \gamma~}

\newcommand{\eqn}[1]{Eq.(\ref{eq:#1})}
\newcommand{\fig}[1]{Fig.(\ref{fig:#1})}

\newcommand{\ukqcdBtoKstargammaresult}{{0.124}^{+20}_{-18}\pm{0.022}}

\newcommand{\ukqcdTtwoqsqmaxresult}{0.269^{+17}_{-9}\pm{0.011}}

\newcommand{\Tonephys}{T_1(q^2{=}0;m_B;m_{K^*})}

\title{Update on the lattice calculation of $\BtoK$ \hfill {\small (hep-lat 9411086)}}
\author{UKQCD collaboration, presented by Brian Gough
\vspace{-2mm}
\address{Physics Department,  University of Southampton, 
Southampton SO17 1BJ, United Kingdom}
}
\begin{document}
\begin{abstract}
We present updated results on the calculation of the matrix elements
for $\BtoK$ in the quenched approximation on a $24^3\times48$ lattice
at $\beta{=}6.2$, using an $O(a)$-improved fermion action.  The
scaling behaviours of the form factors $T_1(q^2{=}0)$ and
$T_2(q^2_{max})$ for the decay are examined and pole model ansatzes
tested. 
\end{abstract}
%
\maketitle
\section{Introduction}
Theoretical interest in the rare decay $B \to K^*\gamma$ as a test of
the Standard Model has been renewed by the experimental results of the
CLEO collaboration \cite{cleo:evidence-for-penguins}. 
The viability of calculating the relevant hadronic matrix elements on
the lattice was first demonstrated by Bernard, Hsieh and
Soni~\cite{bhs:lattice-91} in 1991.

The computational details and results of this work have been described
in references~\cite{ukqcd:penguin-prl} and~\cite{ukqcd:penguin-preprint2}.

\section{Form Factor Definitions}
The hadronic matrix elements can be parametrised by three form factors,
\begin{equation}
\langle K^*(k,\epsilon) | \overline{s} \sigma_{\mu\nu} q^\nu b_R | B(p)
 \rangle  = \sum_{i=1}^3 C^i_\mu T_i(q^2) ,
\end{equation} 
where, 
\begin{eqnarray}
C^{1}_\mu & = & 
2 \varepsilon_{\mu\nu\lambda\rho} \epsilon^\nu p^\lambda k^\rho, \\
C^{2}_\mu & = & 
\epsilon_\mu(m_B^2 - m_{K^*}^2) - \epsilon\cdot q (p+k)_\mu, \\
C^{3}_\mu & = & 
\epsilon\cdot q 
\left( q_\mu - \frac{q^2}{m_B^2-m_{K^*}^2} (p+k)_\mu \right),
\end{eqnarray}
and $q$ is the momentum of the emitted photon.

As the photon emitted is on-shell, the form factors need to be
evaluated at $q^2{=}0$.  In this limit,
\begin{equation}
T_2(q^2{=}0)  = -i T_1(q^2{=}0) ,
\label{eq:T1_T2_equal}
\end{equation} 
and the coefficient of $T_3(q^2{=}0)$ is zero in the on-shell matrix
element.  Hence, the branching ratio can be expressed in terms of a
single form factor, for example $T_1(q^2{=}0)$.

\section{Heavy Quark Scaling}
We calculate with a selection of quark masses near the charm
mass and extrapolate to the $b$-quark scale. In the heavy quark limit,
heavy quark symmetry~\cite{isgur:form-factors} tells us that,
\begin{equation}
\begin{array}{rcl}
T_1(q^2_{max}) &\sim& m_P^{1/2} \\
T_2(q^2_{max}) &\sim& m_P^{-1/2},
\end{array}
\end{equation}
where $m_P$ is the pseudoscalar mass.  Combining this with the
relation $T_2(q^2{=}0) = -iT_1(q^2{=}0)$ constrains the $q^2$
dependence of the form factors. However, it does not provide a scaling
law for $T_1(q^2{=}0)$ without further assumptions about the actual
$q^2$ behaviour of the form factors.

Pole dominance ideas suggest that,
\begin{equation}
T_i(q^2) = {T_i(0)\over (1 - q^2/m_i^2)^{n_i}},
\end{equation}
for $i=1,2$, where $m_i$ is a mass that is equal to $m_P$ plus $1/m_P$
corrections and $n_i$ is a power. Since $1-q^2_{max}/m_i^2 \sim 1/m_P$
for large $m_P$, the combination of heavy quark symmetry and the form
factor relation at $q^2{=}0$ implies that $n_1 = n_2 + 1$. For
example, $T_2(q^2)$ could be a constant and $T_1(q^2)$ a single pole,
or $T_2(q^2)$ could be a single pole and $T_1(q^2)$ a double
pole. These two cases correspond to,
\begin{equation}
T_1(0) \sim \cases{m_P^{-1/2}&single pole\cr
                m_P^{-3/2}&double pole\cr}.
\end{equation}
The data appear visually to favour $T_2(q^2)$ constant
in $q^2$ when $m_P$ is around the charm scale. However, we will
consider both constant and single pole behaviours for $T_2(q^2)$
below.

\section{Results}
As demonstrated in a previous paper~\cite{ukqcd:penguin-prl}, the
evaluation of $T_1(q^2;m_P;m_{K^*})$ is relatively straightforward,
and $T_2$ can be determined in a similar way.
We fit $T_1(q^2)$ to a pole or dipole model in order to obtain the
on-shell form factor $T_1(q^2{=}0)$,
\begin{equation}
T_1(q^2)= {T_1(q^2{=}0) \over 1- q^2/m^2},\quad
{T_1(q^2{=}0) \over (1- q^2/m^2)^2},
\end{equation}
The difference between the two models was found to be negligible.
The form factor $T_2$ was fitted to a pole model or constant 

\begin{figure}
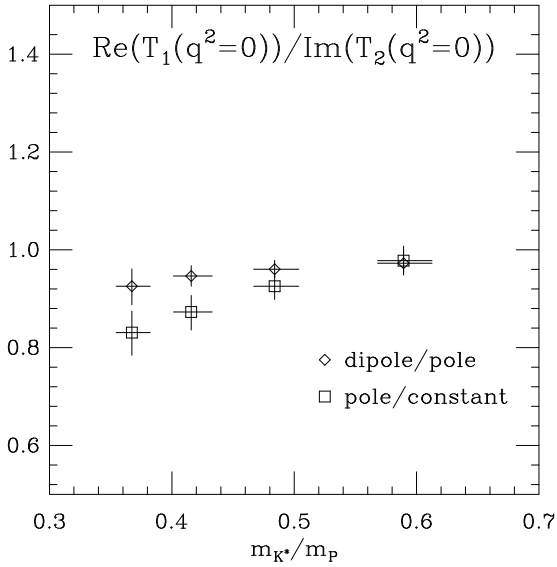

\vspace{9pt}
\pspicture{ratio.epsi}
\caption{The ratio $T_1/T_2$ at $q^2{=}0$ for dipole/pole and pole/constant fits.}
\label{fig:ratio}
\end{figure}

The ratio $T_1/T_2$ at $q^2{=}0$ for dipole/pole and pole/constant
fits is shown in~\fig{ratio}.  The magnitude is found to be consistent
with 1 at low masses, in accordance with the identity
$T_1(0)=iT_2(0)$,~\eqn{T1_T2_equal}. At higher masses, the dipole/pole
fits for $T_1/T_2$ deviate less than the pole/constant fits.
\section{Extrapolation of $T_2(q^2_{max})$ to $m_B$}
\begin{figure}
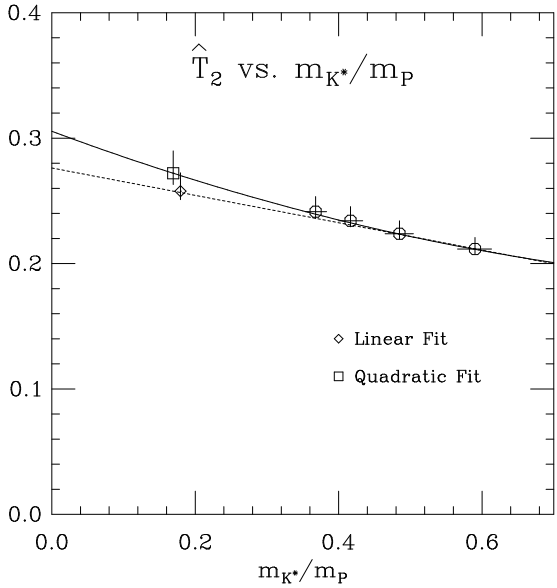

\vspace{9pt}
\pspicture{t2.epsi}
\caption{${\hat T_2}$ extrapolation, with linear and quadratic fits.}
\label{fig:t2-extrapolation}
\end{figure}

In order to test heavy quark scaling, we also extracted the form
factor $T_2$ at maximum recoil, where $q^2=q^2_{max}=(m_P-m_V)^2$, in
the same way as Bernard {\it et al.}~\cite{bhs:penguin-prl}.
In the heavy quark limit, $T_2(q^2_{max})$ is expected to scale as
$m_P^{-1/2}$, analogous to the scaling of $f_B$.  Higher order $1/m_P$
and radiative corrections will also be present.
For convenience, we remove the leading scaling behaviour by forming
the quantity,
\begin{equation}
{\hat T}_2=
T_2(q^2_{max}) 
\sqrt{m_P \over m_B} 
\left({\alpha_s(m_P)\over\alpha_s(m_B)}\right)^{2/\beta_0}.
\end{equation}
The normalisation ensures that ${\hat T}_2=T_2(q^2_{max})$ at the
physical mass $m_B$. Linear and quadratic correlated fits for ${\hat
T_2}$ were carried out with the functions,
\begin{eqnarray}
{\hat T}_2(m_P)&=&A\left(1+{B\over m_P}\right), \\
{\hat T}_2(m_P)&=&A\left(1+{B\over m_P}+{C\over m_P^2}\right),
\end{eqnarray}
and are shown in~\fig{t2-extrapolation}. 
Taking the quadratic fit of $T_2$ at $m_P = m_B$ as the best 
estimate, and the difference between the central values of the
linear and quadratic fits as an estimate of the sytematic error, $T_2$
was found to be, 
\begin{equation}
\label{eq:Ttwo-qsqmax-result}
T_2(q^2_{max};m_B;m_{K^*}) = \ukqcdTtwoqsqmaxresult.
\end{equation}
If the $q^2$ dependence of $T_2$ at $m_B$ were known, this result
could be related to $T_1(q^2{=}0)$ via the identity $T_1(0)=iT_2(0)$.

\section{Extrapolation of $T_1(q^2{=}0)$ to $m_B$}
For $T_1(q^2{=}0)$ we test the two possible scaling laws in the same
way as for $T_2$, by forming the quantity,
\begin{equation}
{\hat T}_1=
T_1(q^2{=}0) 
\left({m_P \over m_B}\right)^n
\left({\alpha_s(m_P)\over\alpha_s(m_B)}\right)^{2/\beta_0},
\end{equation}
where $n=1/2,3/2$.  For $n=3/2$, a similar scaling relationship has
been found using light-cone sum rules by Ali, Braun and
Simma~\cite{ali:3pt-sum-rules}.  The $n=1/2$ case has been suggested
by other sum rules
calculations~\cite{sum-rules}.

\begin{figure}
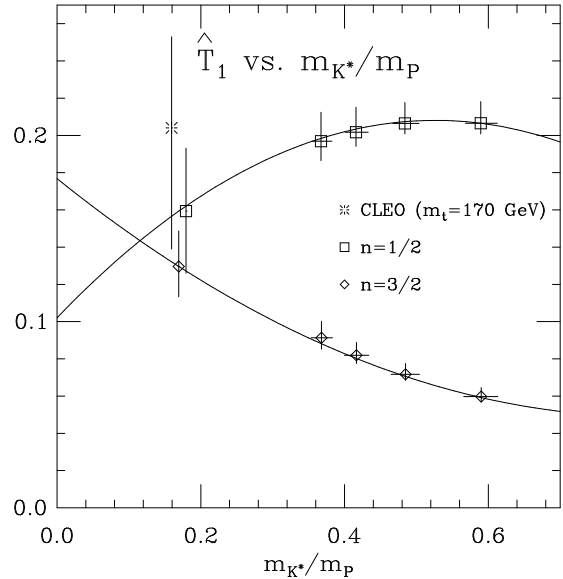

\vspace{9pt}
\pspicture{t1.epsi}
\caption{${\hat T_1}$ extrapolation, for $n=1/2,3/2$ (Points displaced slightly for clarity).}
\label{fig:t1-extrapolation}
\end{figure}

Linear and quadratic fits were carried out with the same functions as
for $T_2$. The two cases $n=1/2,3/2$ are shown
in~\fig{t1-extrapolation}. The $\chi^2/\mbox{d.o.f.}$ are
approximately 1 for the scaling laws, indicating that the models are
statistically valid in the available mass range.

The final results for $\Tonephys$ are taken from the quadratic fit for
$T_1$, with the systematic error estimated as for $T_2$,
\begin{equation}
T_1(q^2{=}0) = \cases{ 0.159^{+34}_{-33}\pm{0.067}&$n=1/2$\cr
                   \ukqcdBtoKstargammaresult.&$n=3/2$\cr}.
\end{equation}
\section{Conclusions}
Further information on the $q^2$ dependence of $T_1$ and $T_2$ is
required to remove the uncertainty in obtaining the form factors at
the physical point $q^2{=}0$, $m_P{=}m_B$.

The authors wish to thank A.~Soni, T.~Bhattacharya and G.~Martinelli
for useful discussions.
\newcommand{\noopsort}[1]{} \newcommand{\printfirst}[2]{#1}
  \newcommand{\singleletter}[1]{#1} \newcommand{\switchargs}[2]{#2#1}


\begin{thebibliography}{10}
\bibitem{cleo:evidence-for-penguins}
{CLEO Collaboration, R. Ammar} {\it et~al.}, Phys. Rev. Lett. {\bf 71},  674
  (1993).

\bibitem{bhs:lattice-91}
C. Bernard, P. Hsieh, and A. Soni, Nucl. Phys. (Proc Suppl.) {\bf B26},
  347  (1992), note that there is a factor of 2 missing in eq.~(4) of this
  paper.

\bibitem{ukqcd:penguin-prl}
{UKQCD Collaboration, K. Bowler} {\it et~al.}, Phys. Rev. Lett. {\bf 72},
  1398  (1994), hep-lat 9311004.

\bibitem{ukqcd:penguin-preprint2}
{UKQCD Collaboration, K. Bowler} {\it et~al.}, hep-lat 9407014  (1994).

\bibitem{isgur:form-factors}
N. Isgur and M. Wise, Phys. Rev. D {\bf 42},  2388  (1990).

\bibitem{bhs:penguin-prl}
C. Bernard, P. Hsieh, and A. Soni, Phys. Rev. Lett. {\bf 72},  1402  (1994),
  hep-lat 9311011.

\bibitem{ali:3pt-sum-rules}
A. Ali, V. Braun, and H. Simma., hep-ph 9401277  (1993).

\bibitem{sum-rules}
P. Ball, hep-ph 9308244 (1993). P. Colangelo, C. Dominguez,
G. Nardulli, and N. Paver, Phys. Lett. B {\bf 317}, 183 (1993), hep-ph
9308264. 
\end{thebibliography}
\end{document}